# Amplitude Modulation Noise Suppression of Dynamic Atom Gravimeters

Wen-Zhang Wang[1,2], Jin-Ting Li[1,2], Dan-Fang Zhang[1,2], Wei-Hao Xu[1,2], Jia-Yi Wei[1,2], Jia-Qi Zhong[1,4], Biao Tang[1], Lin Zhou[1,4], Run-Bing Li[1,3,4], Xi Chen[1,3,*],Jin-Wang[1,3,4,†], Ming-Sheng Zhan[1,3,4,‡]

[1]*Division of Precision Measurement Physics, Wuhan Institute of Physics and Mathematics, Innovation Academy for Precision Measurement Science and Technology, Chinese Academy of Sciences, Wuhan 430071, China*
[2]*School of Physical Sciences, University of Chinese Academy of Sciences, Beijing 100049, China*
[3]*Wuhan Institute of Quantum Technology, Wuhan 430206, China*
[4]*Hefei National Laboratory, Hefei 230088, China*

**ABSTRACT**. Dynamic atom gravimeters enable absolute gravity measurements on moving platforms. However, their performance is severely degraded due to the complex dynamic environment. This paper finds that the amplitude modulation noise (AMN) is a key factor contributing to the degradation of gravity measurement performance. We find that the AMN is induced by the cold atomic cloud trajectory and velocity variation. We build a model to illustrate the principles and magnitude of AMN arising from various experiment processes. Then we propose a method to fit the normalized AMN respect to the kinematic parameters of the cold atomic cloud, and successfully suppress this noise from 0.11 to 0.038 using the fitting result. With this method, we improve the fringe phase resolution from 0.244 rad to 0.092 rad, and reduce the dynamic gravity measurement noise from 2.69 mGal to 1.68 mGal. This study finds and suppresses a key noise source for the dynamic atom gravimeters, which is important for further improving its precision. The proposed method can be also applied for precision enhancement for other dynamic atom interferometer-based sensors, such as the atom gradiometers and gyroscopes.

## I. INTRODUCTION.

Dynamic atom gravimeters overcome the measurement drift of traditional dynamic spring gravimeters [1,2] and enable absolute gravity measurements on moving platforms [3–14]. They play a crucial role in gravity surveying [15,16], resource exploration [17], geophysical studies [18], navigation [19,20], and metrology [21,22]. In dynamic environments, the vibration-induced phase shift of interference fringes typically far exceeds 2π [23,24]. To enable effective gravity measurements, vibration compensation methods have been proposed [23,25]. By synchronizing and matching measurements from the atom interferometer with a classical accelerometer, these methods extend the dynamic measurement range of the atom interferometer, correct for the bias and drift of the classical accelerometer, and thereby achieve continuous, absolute gravity output [26–28].

The principle of vibration compensation involves using the accelerometer's readings to calculate and compensate for the vibration-induced phase shift in the interference fringes. However, under actual dynamic conditions, the cold atomic cloud not only experiences vibration-induced phase shifts but also undergoes changes in velocity and trajectory due to platform motion. These changes lead to variations in efficiency during both the interference and fluorescence detection processes, resulting in fluctuations in the effective amplitude of the interference fringes. This aspect is not accounted for in phase-based vibration compensation schemes. This explains why static atomic gravimeters can achieve interference times on the order of 100 ms and measurement resolutions at the level of 100 μGal/√Hz [29–33], whereas on dynamic platforms, interference times are typically less than 20 ms, and gravity measurement resolution is limited to the mGal/√Hz level [5,8,9]. The amplitude fluctuations of interference fringes caused by the atomic cloud's motion trajectory are a significant factor contributing to these limitations. Analyzing and addressing this issue is key to improve the performance of dynamic atom gravimeters.

This paper investigates the problem from two dimensions: theoretical modeling and analysis of field dynamic measurement data. It focuses on addressing the following questions: 1. Which factors and parameters contribute to amplitude noise in interference fringes, and what is the magnitude of their impact? 2. Can a method be developed to suppress the amplitude noise caused by various factors, and if so, how effective is it? 3. If amplitude noise can be effectively suppressed, will it improve the accuracy of dynamic gravity measurements, and by how much?

The paper is structured around these three questions as follows. In Section II, we introduce the concept of amplitude modulation noise and analyze its magnitude using actual dynamic gravity data. In Section III, we establish a model for the generation of amplitude modulation noise and analyze the five primary mechanisms that produce it during the atom interference and fluorescence detection processes. In Section IV, we propose a simplified quadratic fitting method that successfully correlates amplitude modulation noise with kinematic parameters, and introduce a noise decoupling algorithm to suppress it. In Section V, we compare the fringe fitting results and dynamic gravity measurement outcomes before and after decoupling the amplitude modulation noise. Finally, Section VI presents the conclusions and discussion.

*Contact author: chenxi@apm.ac.cn
†Contact author: wangjin@apm.ac.cn
‡Contact author: mszhan@apm.ac.cn

## II. THE AMPLITUDE MODULATION NOISE IN ATOMIC INTERFERENCE FRINGES

We first introduce the concept of the amplitude modulation noise. For an ideal atom interference fringe, the normalized fluorescence population $P(k)$ can be described by:

$$P(k) = 1 + A \cdot \sin(\phi_k) \quad (1)$$

where $A$ is the fringe amplitude, $\phi_k$ is the fringe phase, and $k$ is the experimental cycle index. During the interference or detection process, $P(k)$ will change due to the interference or detection efficiency change. The proportional change in $P(k)$ relative to the unperturbed case is denoted as $M(k)$. Then the form of $P(k)$ is

$$P(k) = [1 + A \cdot \sin(\phi_k)][1 + M(k)] \quad (2)$$

We refer to $M(k)$ as the normalized amplitude modulation noise (NAMN), and its standard deviation $\sigma_M$ is denoted as the normalized amplitude modulation noise magnitude (NAMNM). Figs. 1(a) and 1(b) show interference fringes for the cases where $M(k)$ is zero and where $M(k)$ following a Gaussian distribution, respectively. When NAMN is present, the interference fringe exhibits broadening and blurring, and its phase measurement precision is reduced. The population histograms of Figs. 1(a) and 1(b) are shown as the shaded areas in Figs. 1(c) and 1(d). For noise-free fringes, its histogram has two sharp peaks at the maximum and minimum population values. For fringe with NAMN, its histogram has lower peaks heights and broader peaks widths.

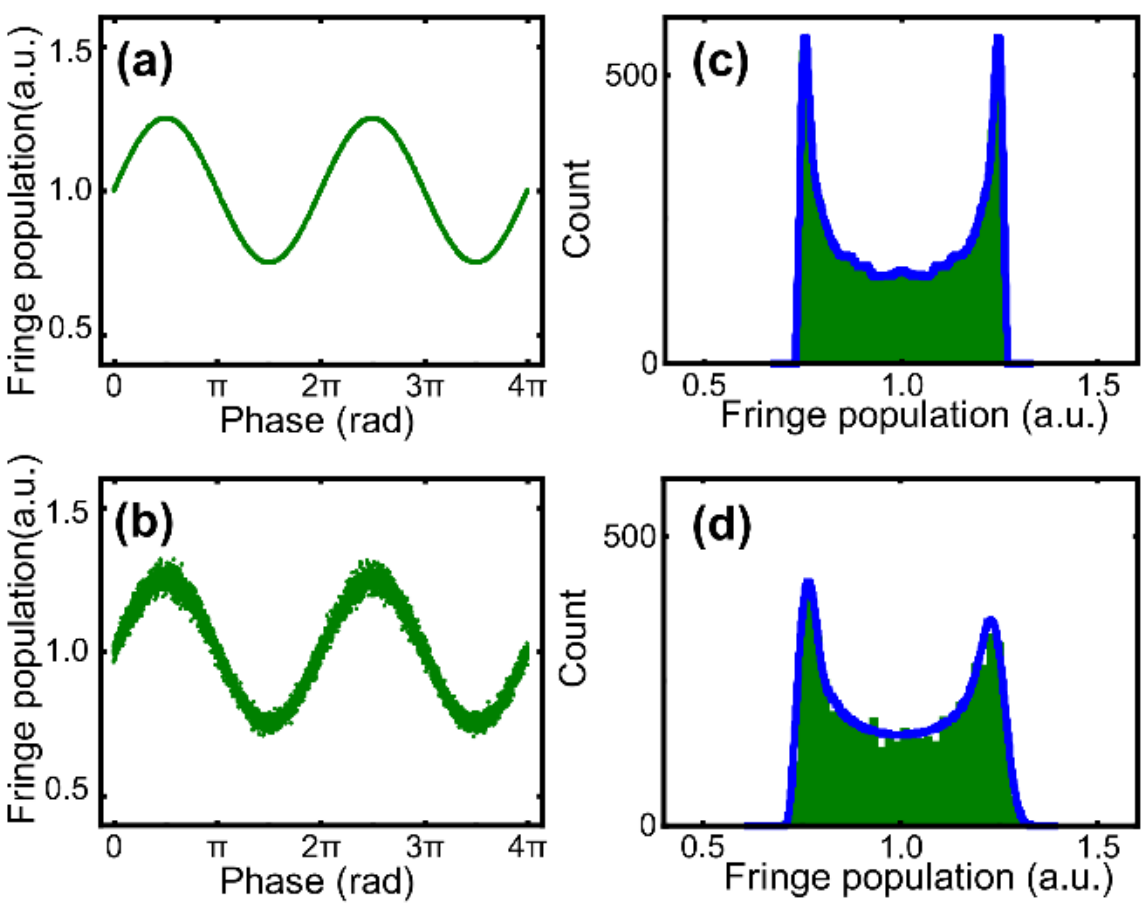

FIG. 1. Simulated data of interference fringes for (a) $A = 0.25$, $\sigma_M = 0$ and (b) $A = 0.25$, $\sigma_M = 0.02$, with their population histograms shown in (c) and (d). The blue lines in (c) and (d) represent the theoretical calculated population curves.

To accurately evaluate the NAMNM from a noisy interference fringe, we propose a method that could estimate the fringe amplitude $A$ and the NAMNM $\sigma_M$ from a noisy data of Sine curve with NAMN. Detailed illustration is in APPENDIX A. Figs. 1(c) and 1(d) show the calculated population curves under the same values of fringe parameters of $A$ and $\sigma_M$, which agree well with the actual histogram, validating the reliability of extracting the NAMN with this method.

We then analyzed the NAMN using field dynamic atom gravity measurement data acquired with a self-developed dynamic atom gravimeter. The schematic diagram of this gravimeter is shown in Fig. 2(a). After cooling, a cloud of $^{85}$Rb atoms is released and initially prepared in the $F$=2 state. Subsequently, three Raman laser pulses interact with the cold atoms, forming a Mach-Zehnder (M-Z) type interference loop. When the cold atom cloud falls into the detection region, as shown in Fig 2(b), the detection lasers excite fluorescence from atoms in the $F$=3 state. An imaging system detects and collects this atomic fluorescence, yielding the interference fringe population $P_{F_3}(k)$. Following this, repump laser is used to transfer atoms from the $F$=2 state back to the $F$=3 state. After a 4 ms interval, the atoms in the $F$=3 state are detected again. This detected population represents the total number of atoms $P_{F_All}(k)$. A classical accelerometer is mounted on the Raman laser mirror to enable hybrid dynamic gravity measurement, as shown in Figs. 2(a), and the principle of vibration phase compensation is introduced in APPENDIX B.

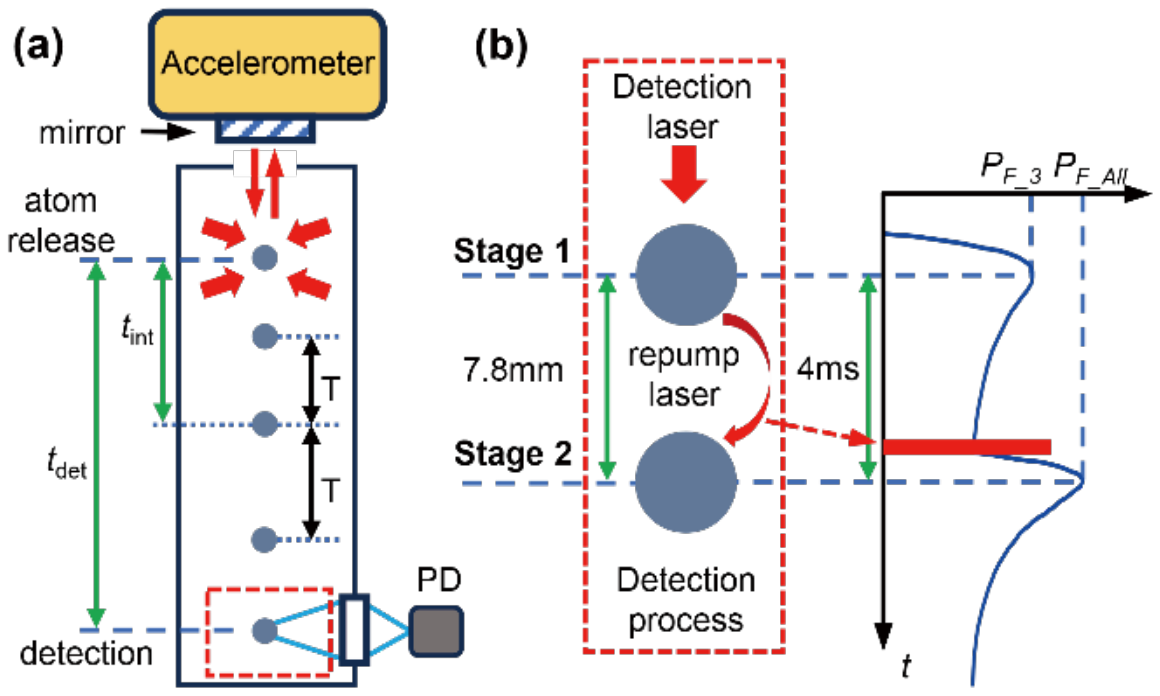

FIG. 2. Schematic overview of the dynamic atom gravimeter. (a) Schematic diagram of the atomic gravimeter. (b) Cold atom cloud in the fluorescence detection region (left panel) and the corresponding fluorescence signal as a function of time (right panel).

*Contact author: chenxi@apm.ac.cn
†Contact author: wangjin@apm.ac.cn
‡Contact author: mszhan@apm.ac.cn

The time interval between the two detection positions is 4 ms, corresponding to a vertical separation of 7.8 mm.

Using this gravimeter, we conducted dynamic gravity measurement experiments under marine conditions during both mooring and along survey lines. The interference time for dynamic gravity measurement was set to T = 15 ms. Figs. 3(a) and 3(b) show typical time-domain acceleration signals under mooring and sailing conditions, where $x$ is the direction of sailing and $z$ is the direction of gravity. Under mooring conditions, the standard deviation of the accelerations in all three axes ($x$, $y$, $z$) are below 0.001 m/s². During sailing, due to wave-induced ship rocking and heaving, the standard deviation of the accelerations increased significantly, reaching 0.01 m/s², 0.03 m/s², 0.08 m/s² in the $x$, $y$ and $z$ directions, respectively. Figs. 3(c) and 3(d) show the typical normalized interference fringes after applying vibration phase compensation for mooring and sailing conditions. It can be seen that under mooring state, the fringes are smooth, whereas under sailing state, the fringes exhibit obvious blurring. The histograms of Figs. 3(c) and 3(d) are shown as Figs. 3(g) and 3(i). The mooring state histogram exhibits the typical bimodal shape of a sine curve, while the sailing state histogram presents a broadened unimodal shape. Since pure phase noise does not alter the histogram shape of a sine curve, the change in the histogram during sailing indicates a significant increase in amplitude-type noise. Figs. 3(e) and 3(f) show normalized total atom number signals for mooring and sailing conditions, with their histograms shown as Figs. 3(h) and 3(j). A similar increase in amplitude noise is observed.

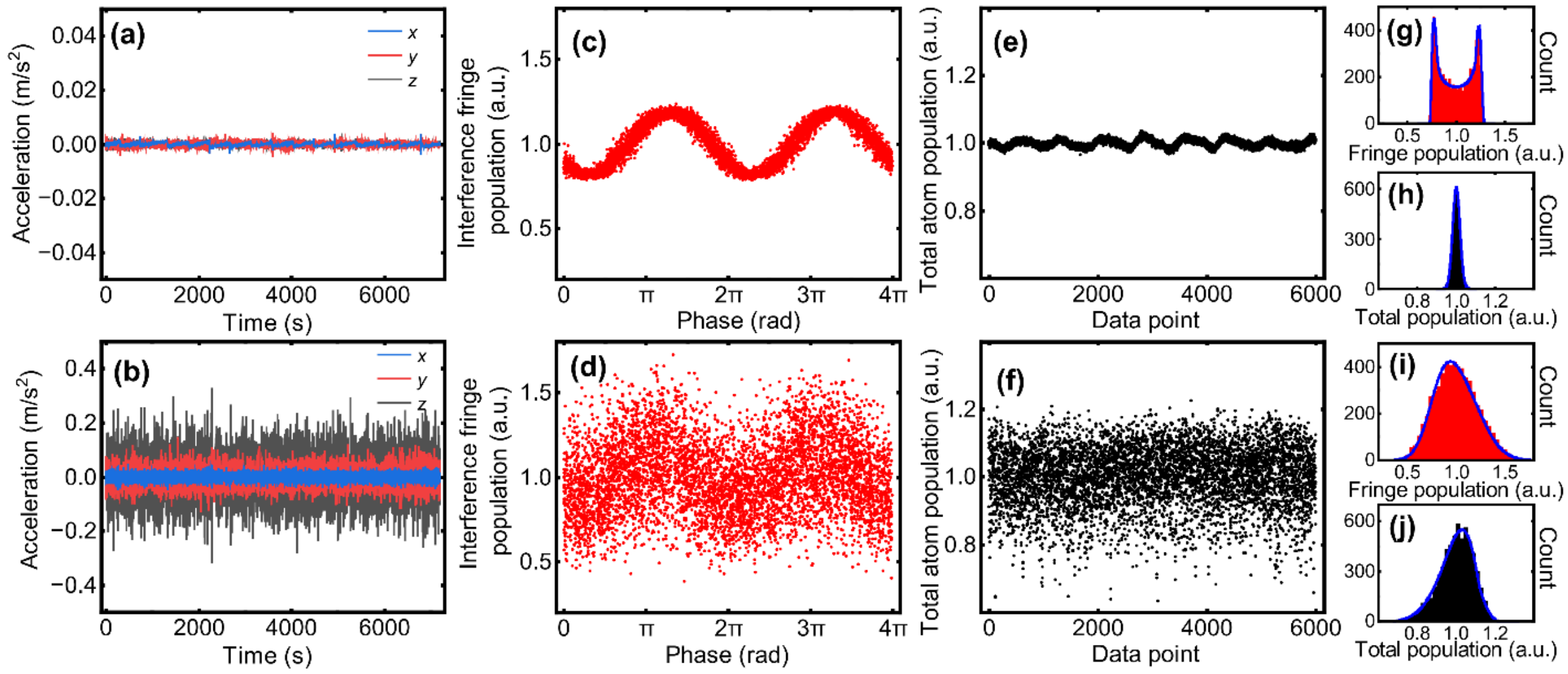

FIG. 3. Measurement signals and their statistical distributions under mooring and sailing conditions. Time-domain acceleration recorded during the mooring (a) and sailing (b) states. Interference fringes after applying vibration phase compensation during the mooring (c) and sailing (d) states, the NAMNM increases from 0.01 to 0.11. Normalized total atom number signals during the mooring (e) and sailing (f) states, the NAMNM increases from 0.011 to 0.088. Histograms of the signals during the mooring (g, h) and sailing (i, j) states.

Using the formula derived in APPENDIX A, we evaluated the fringe amplitude $A$ and the NAMNM $\sigma_M$ for the normalized interference fringes under both mooring and sailing conditions. The theoretical histogram curves obtained after parameter matching are shown as the blue solid lines in Figs. 3(g) and 3(i). For the mooring state, the matched parameters were $A$ = 0.2 and $\sigma_M$ = 0.01. For the sailing state, the matched parameters were $A$ = 0.17 and $\sigma_M$ = 0.11. Compared to the mooring state, the fringe amplitude $A$ did not change much, but the NAMNM increased by approximately a factor of 11. The matched curves show excellent agreement with the histogram of experimental data. This demonstrates that under highly dynamic conditions like sailing, the proposed NAMN becomes a significant factor leading to fringe noise and degradation in gravity measurement performance.

We also compared the fluctuations of the normalized total atom number $P_{F_All}(k)$ between mooring and sailing states. In the mooring state, the standard deviation of $P_{F_All}(k)$ was 0.011. In the sailing state, this standard deviation increased to 0.088.

*Contact author: chenxi@apm.ac.cn
†Contact author: wangjin@apm.ac.cn
‡Contact author: mszhan@apm.ac.cn

Since $P_{F_All}(k)$ is not involved in the interference process, this increasing noise further illustrates the impact of atomic trajectory and velocity on the detection signal noise.

## III. MECHANISMS OF AMPLITUDE MODULATION NOISE

This section provides a quantitative analysis of the mechanisms and magnitudes of NAMN. The cold atom interference fringe is determined by two processes: the atom interference process and the atomic fluorescence detection process. We quantitatively analyze the NAMN arising from each of these processes separately.

During the atom interference process, two primary factors lead to variations in fringe amplitude: Doppler detuning during Raman laser interaction and the horizontal intensity distribution of the Raman laser. First, for the Doppler detuning effect, the induced vertical velocity during the interference process is illustrated in Fig 4(a). This velocity, denoted as $v_{z_\mathrm{int}}(k)$, can be calculated by integrating the $z$ direction acceleration $a_z(k,t)$ as $v_{z_\mathrm{int}}(k) = \int_0^{t_{\mathrm{int}}} a_z(k,t)\,dt$, where $t_{\mathrm{int}}$ = 50 ms is the time from atom release to the interference sequence. This velocity leads to a Doppler detuning of the Raman pulses given by $\delta_{\mathrm{dop_int}}(k) = k_{\mathrm{eff}} v_{z_\mathrm{int}}(k)$, where $k_{\mathrm{eff}}$ is the effective wavevector. To model this effect, we solved the Schrödinger equations involving the atom states and Raman laser pulses. Then we calculated the relationship between the amplitude modulation of the interference fringe and the $v_{z_\mathrm{int}}(k)$. The result is shown in Fig 4(b), which exhibits a Gaussian-like shape. The full width at half maximum (FWHM) of this curve is 0.047 m/s. Second, for the effect of horizontal intensity distribution of the Raman laser, the atomic cloud experiences a horizontal displacement $r_{\rho_\mathrm{int}}(k)$ in a dynamic environment as shown in Fig 4(c). $r_{\rho_\mathrm{int}}(k)$ could be calculated by the horizontal acceleration $a_\rho(k,t)$ as $r_{\rho_\mathrm{int}}(k) = \int_0^{t_{\mathrm{int}}}\int_0^{t} a_\rho(k,t')dt'\,dt$. The Raman laser has an intensity profile in the horizontal plane. $r_{\rho_\mathrm{int}}(k)$ leads to a change in the effective Raman laser intensity, and consequently induces an amplitude modulation of the interference fringe. Using actual experimental parameters, the calculated relationship between the amplitude modulation and $r_{\rho_\mathrm{int}}(k)$ is shown in Fig 4(d), which also follows a Gaussian distribution. Through fitting, its FWHM is found to be 5.59 mm.

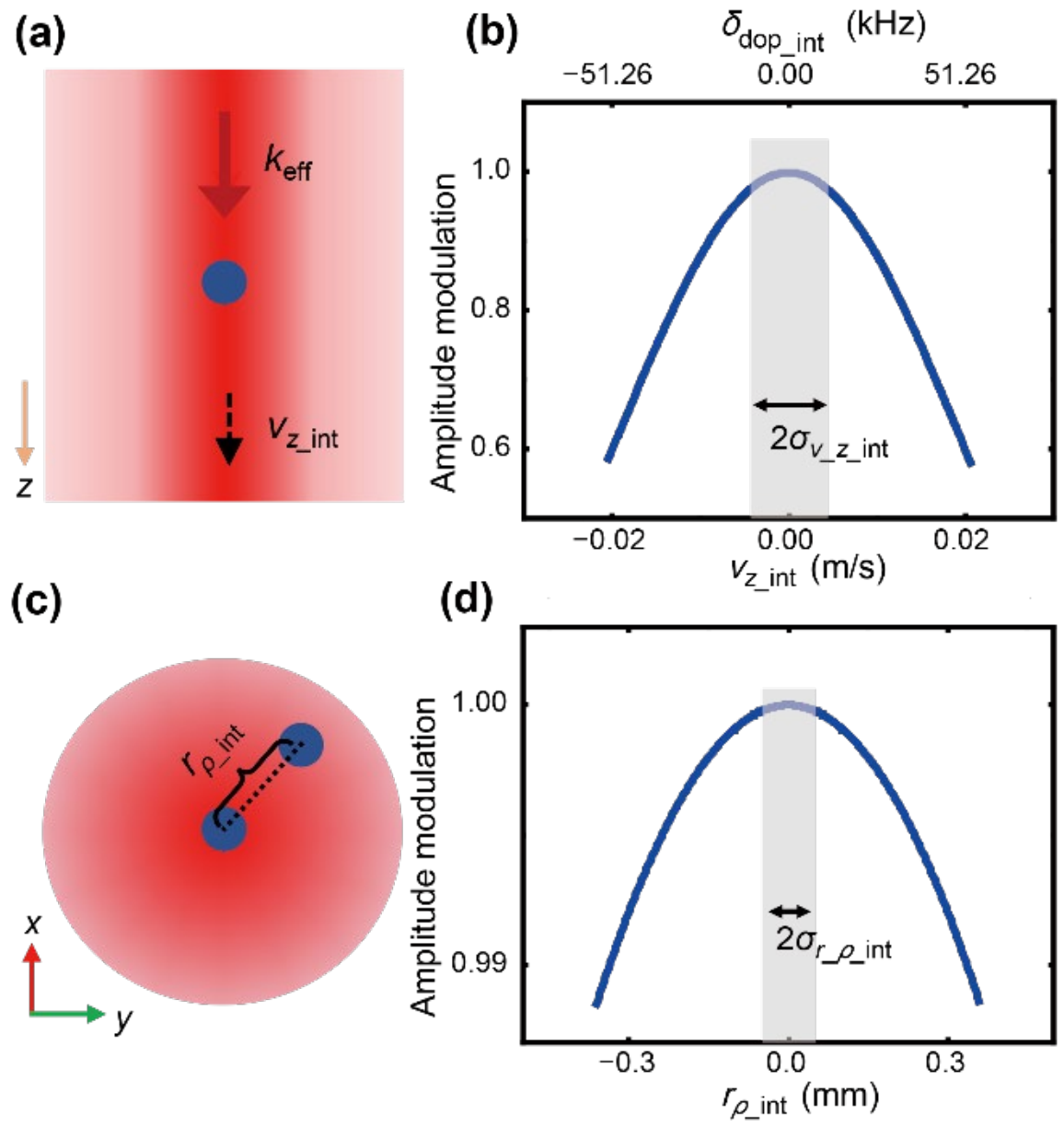

FIG. 4. Mechanisms of amplitude modulation during the interference process. (a) Schematic of Doppler detuning induced by vertical velocity during Raman laser interaction. (b) The relationship between the amplitude modulation and the vertical velocity during the interference process. (c) Schematic of horizontal intensity profile of the Raman laser beam. (d) The relationship between the amplitude modulation and the horizontal displacement during the interference process.

During the fluorescence detection process, the detection efficiency induces an equivalent amplitude modulation of the interference fringe. There are three primary factors that influence the fluorescence detection efficiency: the Doppler detuning of the detection laser, the horizontal intensity distribution of the detection laser, and the vertical position of the atomic cloud during detection. First, for the Doppler detuning effect, the $z$ direction velocity of the atomic cloud $v_{z_\mathrm{det}}(k)$ during detection is calculated as $v_{z_\mathrm{det}}(k) = \int_0^{t_{\mathrm{det}}} a_z(k,t)\,dt$, where $t_{\mathrm{det}}$ = 205 ms is the time from atom release to the fluorescence detection process. This leads to a Doppler shift during detection as $\delta_{\mathrm{dop_det}}(k) = k_{\mathrm{eff}} v_{z_\mathrm{det}}(k)$, thereby affecting the fluorescence detection efficiency as shown in Fig 5(a). We calibrated the relationship between the detection laser frequency and the detection efficiency as shown in Fig 5(b). Its peak region can be approximated by a Gaussian distribution. Through fitting, the corresponding FWHM for $v_{z_\mathrm{det}}(k)$ is 4.33 m/s. Second, for the horizontal laser intensity distribution effect, the horizontal position of the atomic cloud will change due to the dynamic environment, therefore affecting the fluorescence detection efficiency as shown in Fig 5(c). The horizontal position of atoms during detection is calculated as $r_{\rho_\det}(k) = \int_0^{t_{\det}} \int_0^{t} a_\rho(k,t')dt'\,dt$ . Based on the detection laser beam profile, we calculated the relationship between detection efficiency and $r_{\rho_\det}(k)$, as shown in Fig 5(d). The relationship follows a Gaussian distribution with a FWHM of 5.59 mm. Third, for the effect of vertical position of the atomic cloud during detection, the position of the atomic cloud in the $z$ direction is $r_{z_\det}(k) = \int_0^{t_{\det}} \int_0^{t} a_z(k,t')dt'\,dt$ . The excited fluorescence from the atomic cloud is collected by an imaging system as shown in Fig 5(e). Variations in $r_{z_\det}(k)$ lead to changes in detection efficiency. Through calculation, the relationship between $r_{z_\det}$ and detection efficiency is shown in Fig 5(f), which exhibits a Gaussian distribution with a FWHM of 22.1 mm.

*Contact author: chenxi@apm.ac.cn
†Contact author: wangjin@apm.ac.cn
‡Contact author: mszhan@apm.ac.cn

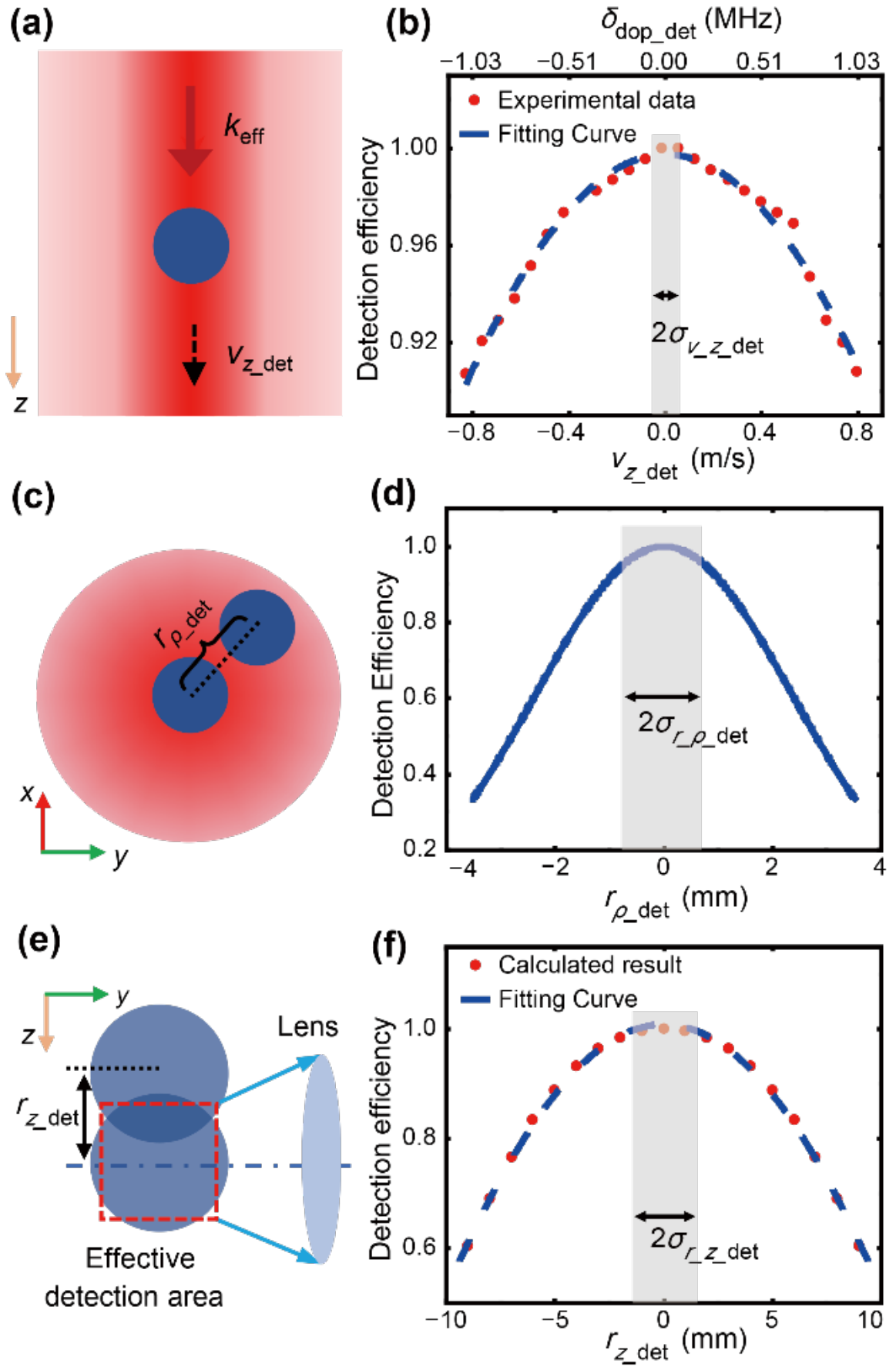

FIG. 5. Mechanisms of amplitude modulation during the fluorescence detection process. (a) Schematic of Doppler detuning induced by vertical velocity during detection laser interaction. (b) The relationship between the detection efficiency and the vertical velocity during the detection process. (c) Schematic of horizontal intensity profile of the detection laser beam. (d) The relationship between the detection efficiency and the horizontal displacement during the detection process. (e) Schematic of the atomic cloud's vertical position during the detection process. (f) The relationship between the detection efficiency and the vertical displacement during the detection process.

We use the above models to estimate the amplitude modulation under sailing conditions. All the factors that induce the amplitude modulation are related to the kinematic parameters of the atomic cloud during the interference and fluorescence detection processes, such as $v_{i_\mathrm{int}}$ and $r_{i_\det}$ (with $i = \rho,\ z$). These parameters can be calculated from the acceleration values $a_i$ during the cold atom interference process. By using acceleration data and the interference time sequence, we calculated the standard deviation of the kinematic parameters of the above factors under sailing conditions, as shown by the shaded areas in Figs. 4(b), 4(d) and Figs. 5(b), 5(d), 5(f), and listed in Table I. The fluctuation of these kinematic parameters will induce amplitude modulation noise, and we derive the formula of the magnitude of NAMN for a Gaussian distribution. Detailed derivation is shown in APPENDIX C.

The calculated NAMNM values for each factor are listed in Table I. The largest contribution arises from the factor of 0.056 during the detection process. Then, we calculated the total NAMNMs for the interference fringe $P_{F_3}(k)$ and for the total atom number $P_{F_All}(k)$. For the former one, we sum over all the five factors, and for the latter one, we sum over the three factors during the fluorescence detection process. The estimated values are 0.077 and 0.061, respectively. Compared with the actual experimental results of 0.11 and 0.088 obtained in Section II, they show agreement in order of magnitude. In this calculation, we assumed the mean values of all kinematic parameters were at the center of their Gaussian distributions and did not consider the position difference of $r_{z_\det}$ between the measurements of $P_{F_3}(k)$ and $P_{F_All}(k)$ during the detection process.

TABLE I: NAMNM values calculated from theoretical models.

| Measurement sequence | Physical mechanism | FWHM | Standard deviation of kinematic parameters | Calculated NAMNM |
| --- | --- | --- | --- | --- |

*Contact author: chenxi@apm.ac.cn
†Contact author: wangjin@apm.ac.cn
‡Contact author: mszhan@apm.ac.cn

| Interference process | Doppler detuning | 0.047 m/s | $\sigma_{v_z_\mathrm{int}}$ = 0.004 m/s | 0.028 |
|---|---|---|---|---|
| | Laser intensity profile | 5.59 mm | $\sigma_{r_\rho_\mathrm{int}}$ = 0.041 mm | 0.002 |
| Detection process | Doppler detuning | 4.33 m/s | $\sigma_{v_z_\mathrm{det}}$ = 0.016 m/s | 0.00005 |
| | Laser intensity profile | 5.59 mm | $\sigma_{r_\rho_\mathrm{det}}$ = 0.67 mm | 0.056 |
| | Vertical position shifts | 22.1 mm | $\sigma_{r_z_\mathrm{det}}$ = 1.68 mm | 0.023 |
| NAMNM of the interference fringe $P_{F_3}(k)$ | | | | 0.077 |
| NAMNM of the total atom number $P_{F_All}(k)$ | | | | 0.061 |

## IV. FITTING AND DEDUCTION OF THE AMPLITUDE MODULATION NOISE

The models developed in Section III give the formal expressions of NAMN in terms of kinematic parameters. However, they do not provide the exact values of key parameters such as the width and the centers of the Gaussian curves. Fitting to actual experimental data is required to determine these parameters. There are two challenges for the fitting process. First, the relationship between NAMN and kinematic parameters is described by Gaussian curves, as illustrated in Section III. The superposition of these functions results in a multimodal distribution. It is difficult to obtain accurate fitting results when the widths and centers of these peaks are unknown. Second, NAMN depends on multiple kinematic parameters. The resulting high-dimensional fitting space makes the fitting process both complex and inefficient.

To address these issues, we introduce two approximations for the form of the NAMN. First, the distributions of the kinematic parameters are concentrated near the centers of the Gaussian functions, as shown in Figs. 4 and 5. We therefore approximate the Gaussian functions with quadratic functions for simplicity. The superposition of several quadratic functions remains a quadratic function, thereby avoiding the complex form of a multimodal Gaussian mixture. Second, under actual marine sailing conditions, the dominant motion frequency is very low, with a period of about 10 s. The time from atom release to fluorescence detection in our instrument is about 0.2 s. During this short interval, the acceleration can be treated as constant. The kinematic parameters in the interference and fluorescence detection process, such as $v_{i_\mathrm{int}}$, $r_{i_\mathrm{det}}$, are proportional to $a_i$. Thus, the various kinematic parameters can be reduced to three independent parameters in the three spatial dimensions, which could be $a_i$, $v_i$ or $r_i$. Here, we set these three parameters as $r_{i_\mathrm{det}}$, since the dominant contributions of NAMN are induced by $r_{i_\mathrm{det}}$ as illustrated in Table I.

The simplified formula for NAMN is then

$$F(k) = \sum_i \alpha_i \left(r_{i_{\mathrm{det}}}(k) - r_{i0}\right)^2 + \sum_{i\neq j} \beta_{i,j} r_{i_\mathrm{det}}(k) r_{j_\mathrm{det}}(k) + F_0 \qquad (3)$$

where $i$, $j$ represents $x$, $y$ and $z$, $\alpha_i$ is the quadratic coefficient for each individual axis, $r_{i0}$ is the corresponding center position. $\beta_{i,j}$ are the cross-coupling quadratic coefficients between different axes, and $F_0$ is the intercept. Since the acceleration in the $x$ direction is much smaller than the other two axes, we omit the terms that related to $r_{x_\mathrm{det}}$ in Eq. (3) for further simplification.

We first perform a two-dimensional fit of the normalized total atom number $P_{F_All}(k)$ using Eq. (3). As shown in Fig. 6(a), $P_{F_All}(k)$ exhibits a regular surface shape as a function of the displacements $r_{i_\mathrm{det}}$. The fitting surface is denoted as $F_{P_Fall}(k)$, is obtained through this procedure. The corresponding fit parameters are listed in Table II. To better illustrate the agreement between the fitted surface and the experimental data, we extract two cross-sections' curves from the fitted surface by setting $r_{y_\mathrm{det}} = 0$ and $r_{z_\mathrm{det}} = 0$, and compare these two curves with actual experimental data of $P_{F_All}(k)$, as shown in Figs. 6(b) and 6(c). The curves show good agreement with the experimental data. We then suppress the NAMN by dividing the experimental data of $P_{F_All}(k)$ by the fitted surface $F_{P_Fall}(k)$, obtaining the corrected signal $R_{P_Fall}(k) = P_{F_All}(k)/F_{P_Fall}(k)$. Fig 6(d) and 6(e) shows the original experiment data $P_{F_All}(k)$ and the processed data $R_{P_Fall}(k)$. Their corresponding histograms are shown in Figs. 6(f) and 6(g), respectively. The NAMNM is reduced from 0.088 to 0.021, with a noise rejection ratio of 4. The NAMNM of the processed signal $R_{P_Fall}(k)$ in sailing state is close to that of the origin signal $P_{F_All}(k)$ in mooring state, which is 0.011.

*Contact author: chenxi@apm.ac.cn
†Contact author: wangjin@apm.ac.cn
‡Contact author: mszhan@apm.ac.cn

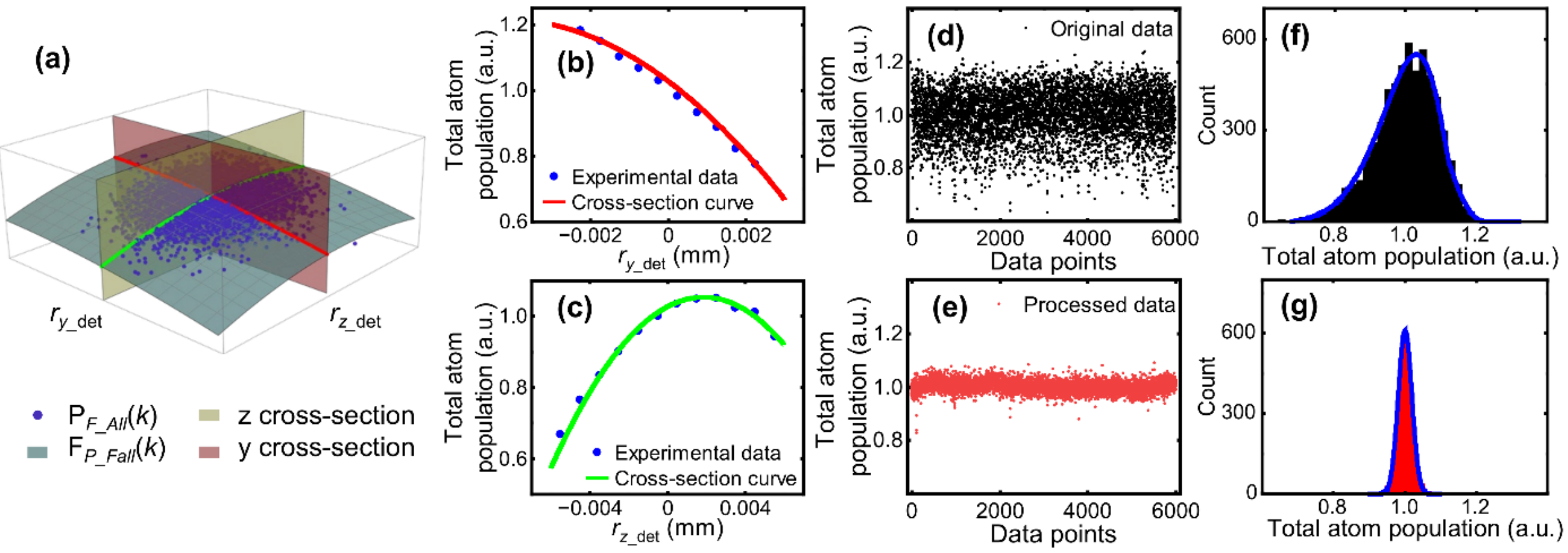

FIG. 6. Fitting of the normalized total atom number $P_{F_All}(k)$ and suppression of NAMN. The NAMNM decreases from 0.088 to 0.021 using the proposed method. (a) $P_{F_All}(k)$ and the 2D fitted surface $F_{P_Fall}(k)$. (b) Cross-section at $r_{z_\mathrm{det}} = 0$ versus experimental data. (c) Cross-section at $r_{y_\mathrm{det}} = 0$ versus experimental data. (d) and (e) Original signal $P_{F_All}(k)$ and the processed signal $R_{P_Fall}(k)$. (e) and (f) Histogram of $P_{F_All}(k)$ and $R_{P_Fall}(k)$.

Then we processed the interference fringe data $P_{F_3}(k)$ similarly to the above procedure. The resulting 2D fitted surface $F_{P_F3}(k)$ and the corresponding cross-sections' curves are shown in Fig. 7(a), Figs. 7(b) and 7(c). The fitted curves exhibit good agreement with the experimental data. Then we suppress the NAMN by $R_{P_F3}(k) = P_{F_3}(k)/F_{P_F3}(k)$. The original experimental data $P_{F_3}(k)$ and the processed data $R_{P_F3}(k)$ are shown in Fig. 7(d) and 7(e). The fringe obtained from $R_{P_F3}(k)$ is significantly more distinct than that from $P_{F_3}(k)$, indicating a reduction in NAMN. More direct evidence is provided by their histograms, as shown in Figs. 7(f) and 7(g). The histogram evolves from a single-peak shape to a distinct bimodal distribution. Using the theoretical histogram formula in APPENDIX A , we evaluated the fringe amplitude $A$ and NAMNM $\sigma_M$. The results are $A$ = 0.17 and $\sigma_M$ = 0.11 for $P_{F_3}(k)$, and $A$ = 0.17 and $\sigma_M$ = 0.038 for $R_{P_F3}(k)$. This corresponds to a reduction factor of 2.9 in the NAMNM.

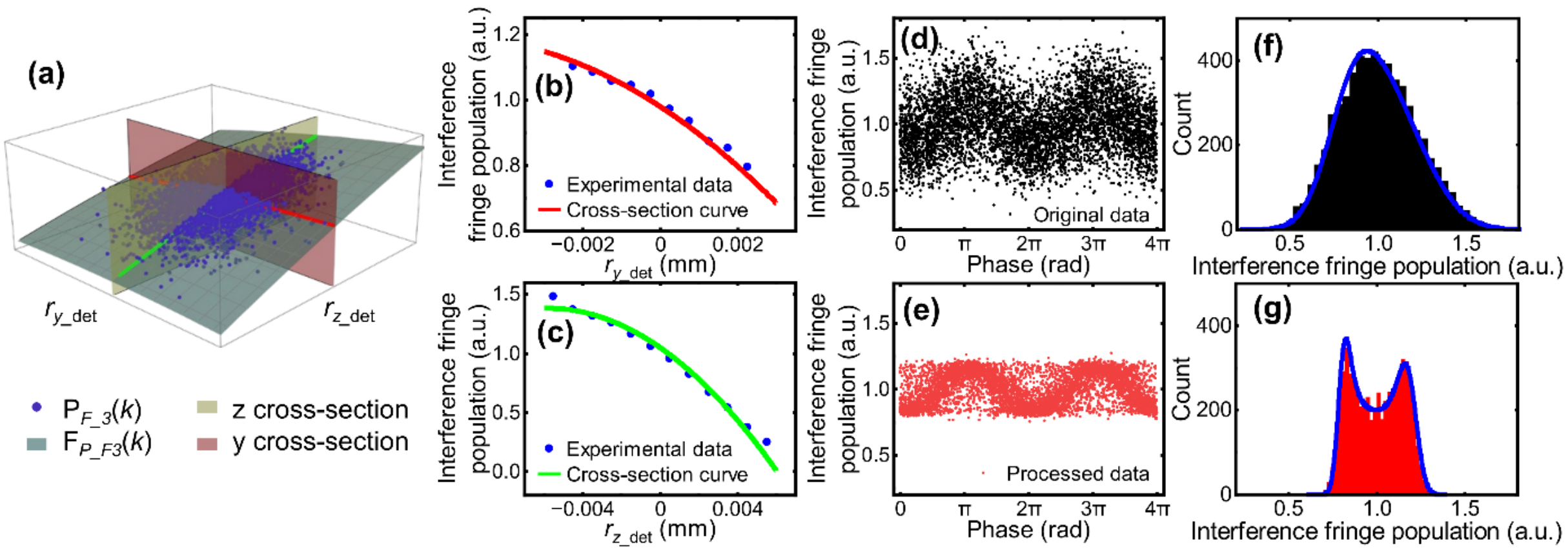

FIG. 7. Fitting of the interference fringe data $P_{F_3}(k)$ and suppression of NAMN. The NAMNM decreases from 0.11 to 0.038 using the proposed method. (a) $P_{F_3}(k)$ and the 2D fitted surface $F_{P_F3}(k)$. (b) Cross-section at $r_{z_\mathrm{det}} = 0$ versus experimental data. (c) Cross-section at $r_{y_\mathrm{det}} = 0$ versus experimental data. (d) and (e) Original signal $P_{F_3}(k)$ and the processed signal $R_{P_F3}(k)$. (e) and (f) Histogram of $P_{F_3}(k)$ and $R_{P_F3}(k)$.

Table II summarizes the fitting results for the parameters of $P_{F_All}(k)$ and $P_{F_3}(k)$. Several key features can be observed from these results. First, the values of coefficients $\alpha_i$ are similar for $P_{F_All}(k)$ and

*Contact author: chenxi@apm.ac.cn
†Contact author: wangjin@apm.ac.cn
‡Contact author: mszhan@apm.ac.cn

$P_{F_3}(k)$ . This indicates that they share same mechanism for both signals. The values of $\alpha_i$ are much larger than those of the coupling coefficient $\beta_{i,j}$. This indicates that the dominant contribution to NAMN originates from independent axes, while the coupling between the different directions is relatively weak. Second, the fitted center positions $r_{y0}$ are in good agreement for both signals. The primary mechanism of NAMN induced by $r_{y_\mathrm{det}}(k)$ is the horizontal distribution of the detection laser beam, as illustrated in Table I. Since $r_{y_\mathrm{det}}(k)$ is similar for the two atom clouds ($F$=3 and total atom number). The agreement in $r_{y0}$ provides validation for the model developed in Section III. Third, the fitted center positions $r_{z0}$ exhibit a clear offset of 7.7 mm between the two signals. This could be explained by the detection time difference of $P_{F_All}(k)$ and $P_{F_3}(k)$, as shown in Fig. 2(c). The detection times for $P_{F_All}(k)$ and $P_{F_3}(k)$ are separated by 4 ms, the calculated position difference due to gravity is approximately 7.8 mm, which agrees well with the fitting result, further verifying the validation of the model analysis.

Table II. Fitting results of the parameters of Eq. 3 with signal of $P_{F_All}(k)$ and $P_{F_3}(k)$.

| | $\alpha_y$ | $\alpha_z$ | $r_{y0}$ | $r_{z0}$ | $\beta_{y,z}$ | $F_0$ |
|---|---|---|---|---|---|---|
| $P_{F_All}(k)$ | $-7028\ \mathrm{m}^{-2}$ | $-9835\ \mathrm{m}^{-2}$ | −4.5 mm | 1.9 mm | $-1563\ \mathrm{m}^{-2}$ | 1.52 |
| $P_{F_3}(k)$ | $-9452\ \mathrm{m}^{-2}$ | $-7773\ \mathrm{m}^{-2}$ | −4.3 mm | −5.8 mm | $-1595\ \mathrm{m}^{-2}$ | 1.24 |

## V. DYNAMIC GRAVITY MEASUREMENT RESOLUTION IMPROVEMENT

To show the effectiveness of the NAMN suppression, we analyzed gravity measurement data from three survey lines, and compared the fitting phase uncertainty $\sigma_\phi$ of the interference fringes and the gravity measurement values before and after the noise suppression. Each line is 45 km long and a survey duration of about 120 min. For each line, the data are divided into 30 interference fringes. The fitting phase uncertainties of the fringes are shown in Fig 8(a), the averaged values of $\sigma_\phi$ for the three survey lines before and after the NAMN suppression are 0.244 rad and 0.092 rad. which shows a resolution improvement of 2.7.

The measured gravity values from dynamic atom gravimeter were compared with those from a strapdown gravimeter installed on the same vessel to calculate the gravity difference $\Delta g$. The resulting $\Delta g$ before and after NAMN suppression are shown in Fig. 8(b). The standard deviation of $\Delta g$ for the three survey lines before and after NAMN suppression are 2.69 mGal and 1.68 mGal, achieving a noise reduction factor of 1.6.

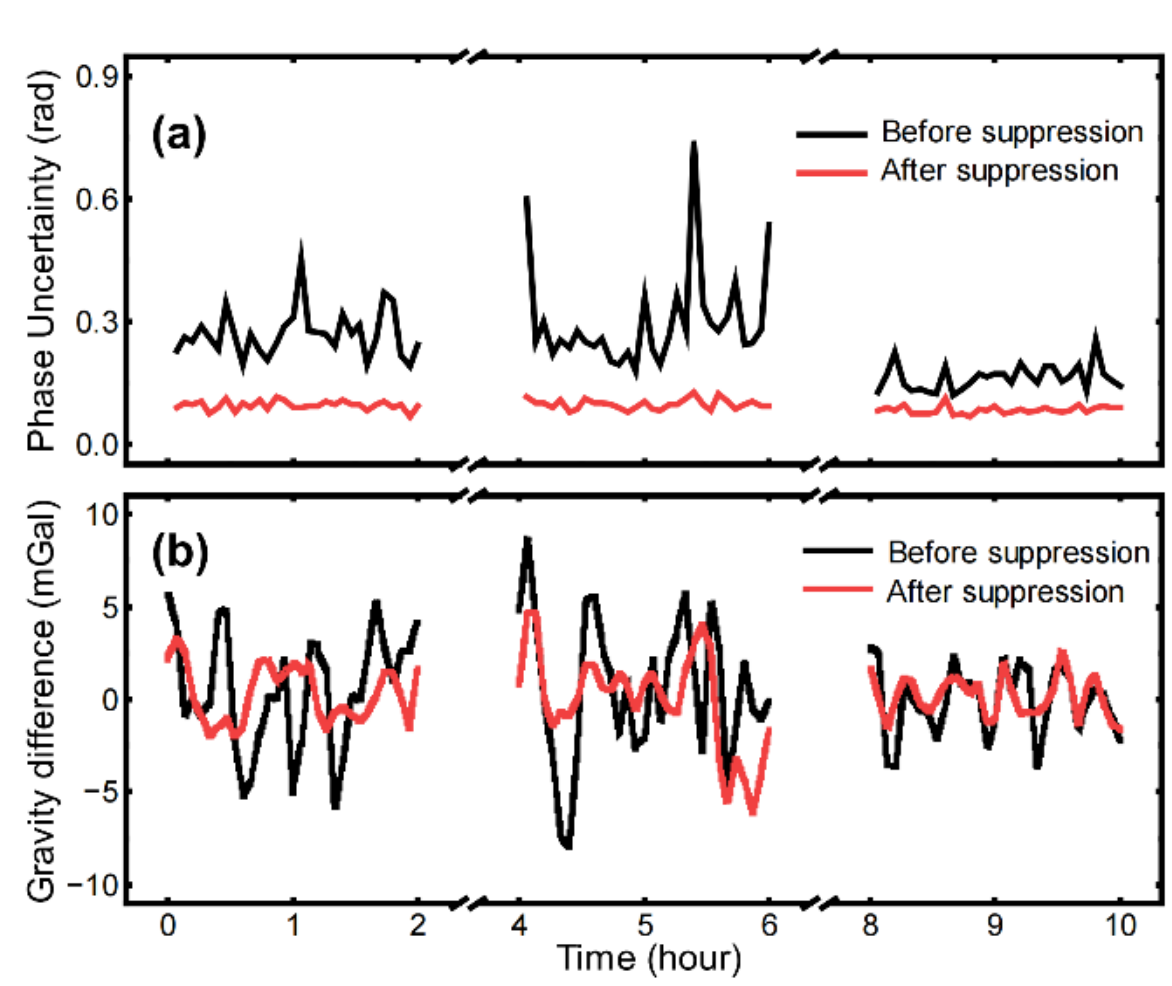

FIG. 8. Performance improvement after NAMN suppression (a)Fitting phase uncertainty of the interference fringes before and after NAMN suppression. (b) Gravity difference before and after NAMN suppression.

## VI. CONCLUSION AND DISCUSSION

This paper reports a systematic study of interference fringe amplitude noise under dynamic conditions. We first introduce the concept of normalized amplitude modulation noise (NAMN), and establish a quantitative evaluation method to extract NAMN from sinusoidal interference fringes. Then we systematically analyze the physical mechanisms that generate NAMN. Based on this analysis, we propose a quadratic fitting method to evaluate and suppression this noise. This method could effectively reduce the NAMN, enhance the fringe phase resolution and reduce the dynamic gravity measurement noise.

*Contact author: chenxi@apm.ac.cn
†Contact author: wangjin@apm.ac.cn
‡Contact author: mszhan@apm.ac.cn

Future studies could focus on NAMN suppression through the improvement of experimental parameters and the optimization of the suppression method. For the experimental parameters, Increasing the laser beam size and shortening the atomic free-fall time could reduce the resulted NAMN. For the method optimization. General motion conditions with high-frequency acceleration noise (such as in airborne gravity measurements) should be considered, where the proportional relationships between acceleration and other kinematic parameters are no longer valid. A more general fitting method should be considered.

This study provides a systematic investigation of amplitude noise in dynamic gravity measurements and proposes an effective suppression method. It can not only improve the accuracy of dynamic gravimetry but also provides important guidance for the precision enhancement of atomic gravity gradiometers and atom gyroscopes on moving platforms.

## ACKNOWLEDGMENTS

This work was supported by the Space Application System of China Manned Space Program (Second batch of the Scientific Experiment Project, JC2-0576), the Innovation Program for Quantum Science and Technology (2021ZD0300603, 2021ZD0300604), the Hubei Provincial Science and Technology Major Project (ZDZX2022000001), the Defense Industrial Technology Development Program (JCKY2022130C012), the National Natural Science Foundation of China (U25D8014, 12174403, W2412045, 12204493, 12304569), the Natural Science Foundation of Hubei Province (2022CFA096), the Wuhan Dawn Plan Project (2023010201020282), The China Postdoctoral Science Foundation (2020M672453).

## APPENDIX A: HISTOGRAM-BASED METHOD FOR NAMNM ESTIMATION

The NAMNM is difficult to extract from a noisy interference fringe using the sine fitting method. This is because that both the phase noise and NAMN will induce the fitting phase error. The fitting method is hard to distinguish these two kinds of noise. Here, we propose a method to estimate NAMN. The basic principle is as following. First, we calculate the theoretical cumulative distribution function (CDF) from a sine interference curve with amplitude of $A$ and NAMNM of $\sigma$. Then the theoretical bin counts for a given histogram are obtained from the differences of the CDF at the bin edges, multiplied by the total number of samples $N$. Finally, we compare the theoretical bin counts with the experimental histogram. By find the minimize difference of these two functions, we obtain the optimize value of $A$ and $\sigma$.

*Contact author: chenxi@apm.ac.cn
†Contact author: wangjin@apm.ac.cn
‡Contact author: mszhan@apm.ac.cn

Assume the noisy fringe has the following form:

$$y(k) = [1 + A \cdot \sin(\phi_k)] \cdot \delta P(k), k = 1,2, \ldots, N, \tag{A1}$$

where $\phi_k$ is the phase, and is uniformly distributed over (0, 2π), $A$ is the fringe amplitude, and $\delta P(k) \sim \mathcal{N}(\mu = 1, \sigma^2)$ is a Gaussian noise with mean value of 1 and standard derivation of $\sigma$.

For a given phase $\phi$, the conditional CDF of $y$ is given by

$$F(y|\phi) = \Phi\left(\frac{y-(1+Asin\phi)}{\sigma(1+Asin\phi)}\right), \tag{A2}$$

where $\Phi(x) = \int_{-\infty}^{x} \frac{1}{\sqrt{2\pi}} e^{-\frac{t^2}{2}} dt$ is the standard normal CDF. Averaging over the uniformly distributed phase $\phi$ yields the averaged CDF of $y$ as

$$F(y) = \frac{1}{2\pi} \int_0^{2\pi} \Phi\left(\frac{y-(1+Asin\phi)}{\sigma(1+Asin\phi)}\right) d\phi, \tag{A3}$$

with finite number of $\phi$, we replace the integral of $\phi$ with a discrete average of $\phi_k$ as

$$F(y) = \frac{1}{N} \sum_{k=1}^{N} \Phi\left(\frac{y-(1+Asin\phi_k)}{\sigma(1+Asin\phi_k)}\right), \tag{A4}$$

where $k$ is the index of the phase sampling points.

To compare with the measured histogram, the number $N$ in the calculation is set the same as the number of fringe's data point in the experiment. The value range and number of bins for the measured histogram are $R_y$ and $M$. $R_y$ is set that it covers 10% more than the actual range of $y$ on each side. $M$ is set by the Terrell–Scott rule [34], which has a value of $M \approx \sqrt[3]{2N} \times (1 + 20\%)$ [35]. Then the theoretical calculated count for a bin in the range of $(y_j, y_{j+1})$ is

$$\hat{n}_j = N \cdot [F(y_{j+1}) - F(y_j)], \tag{A5}$$

where $j$=0,1,…,$M$ and the bin width $\Delta y = y_{j+1} - y_j = R_y/M$.

The squared error of the calculated count $\hat{n}_j$ and the experiment measured count $n_j$ is

$$S = \sum_{j=1}^{M} [\hat{n}_j - n_j]^2. \tag{A6}$$

We find the values $A$ and $\sigma$ that minimize the value of $S$. The optimize value of $\sigma$ is just the NAMNM that we want to measure.

## APPENDIX B: PRINCIPLE OF VIBRATION INDUCED PHASE COMPENSATION

The dynamic hybrid gravity-measurement principle is illustrated in Fig. A1. A detailed description of this principle can be found in Ref. [9].

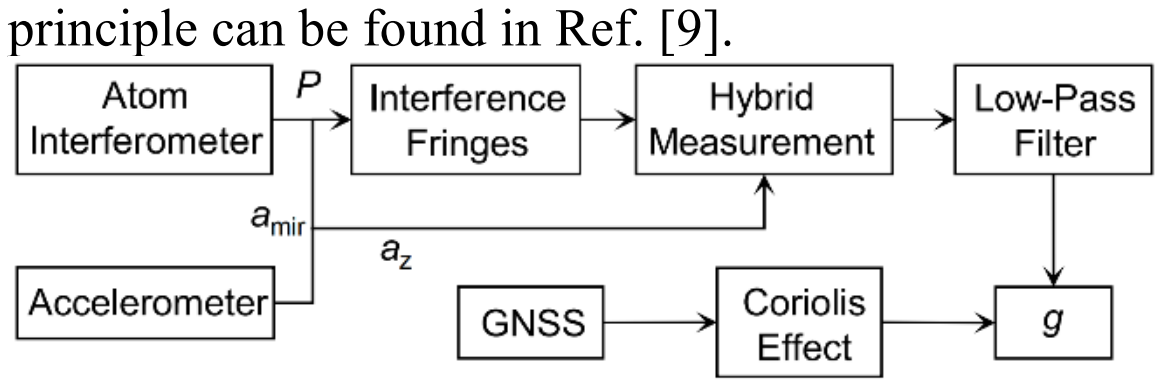

Fig. A1. Schematic of the AI-based dynamic gravity-measurement principle.

A key technique in hybrid gravity measurement is the vibration-induced phase compensation. This phase compensation is based on a hybrid measurement architecture combining an atom interferometer and a classical accelerometer, as shown in Fig. A1. In dynamic environments, platform vibrations introduce additional phase noise via the Raman mirror, resulting in blurred interference fringes. The total phase of the interferometer during the interference process can be expressed as:

$$\phi = (k_{\text{eff}} g - 2\pi\alpha)\mathrm{T}^2 + \Delta\phi_{\text{vib}}, \quad \text{(B1)}$$

where $k_{\text{eff}}$ is the effective wave vector of the Raman laser, $g$ is the gravitational acceleration, $\alpha$ is the laser frequency chirp rate, T is the pulse interval time, and $\Delta\phi_{\text{vib}}$ is the vibration-induced phase shift. The key of the vibration compensation method is real-time measurement and subtraction of this phase shift [4,36–41]. Specifically, a classical accelerometer mounted near the mirror measures the vibration acceleration of the mirror $a_{\text{mir}}$ in real time. The vibration-induced phase shift is then calculated using the acceleration sensitivity function $g_a(t)$ of the atom interferometer [42]:

$$\Delta\phi_{\text{vib}} = k_{\text{eff}} \int_{-\infty}^{\infty} g_a(t) a_{\text{mir}}(t)\, dt. \quad \text{(B2)}$$

Considering the mechanical transfer characteristics between the accelerometer output and the actual mirror vibration, a transfer function model is used to convert the raw accelerometer output $a_z(t)$ into the true mirror acceleration $a_{\text{mir}}(t)$ [43]. Finally, $\Delta\phi_{\text{vib}}$ is subtracted from the total phase to obtain the compensated interference phase, from which the gravitational acceleration is extracted.

## APPENDIX C: NAMNM ESTIMATION FOR A GAUSSIAN FUNCTION

Here we derive the NAMNM when the NAMN $y$ follows a Gaussian distribution with respect to a kinematic parameter $x$. The Gaussian distribution is

$$y = \frac{A}{w\sqrt{\pi/2}} e^{-2\frac{x^2}{w^2}}. \quad \text{(C1)}$$

Assume there are $n$ random $x$ with zero mean and standard deviation $\sigma_n$, and $\sigma_n$ is much smaller than the width of the Gaussian distribution, i.e. $\sigma_n \ll w$. Then the result maximum value of $y$ is $y_0 = \frac{A}{w\sqrt{\pi/2}}$, and $y$ exhibits a random distribution less than $y_0$. Near the peak, the Gaussian function can be approximated by a quadratic function. Expanding $y$ around $x = 0$ gives $y \approx y_0(1 - \frac{2x^2}{w^2})$ . The variance ${\sigma_y}^2$ of $y$ can be calculated as

$${\sigma_y}^2 = E[(y - E[y])^2] = E\left[\left(y_0\left(1 - \frac{2x^2}{w^2}\right) - y_0\left(1 - \frac{2}{w^2}E[x^2]\right)\right)^2\right] = (\frac{2y_0}{w^2})^2 E[(x^2 - E[x^2])^2]. \quad \text{(C2)}$$

For a Gaussian distribution of $x$ , $E[x^4] = 3\sigma_n^4$ , $E[x^2] = \sigma_n^2$, so we have:

$$E[(x^2 - \sigma_n^2)^2] = E[x^4] - 2\sigma_n^2 E[x^2] + \sigma_n^4 = 2\sigma_n^4. \quad \text{(C3)}$$

Substitute Eq. (C3) into Eq. (C2) yields

$${\sigma_y}^2 = (\frac{2y_0}{w^2})^2 \cdot 2\sigma_n^4. \quad \text{(C4)}$$

By substituting the form of $y_0$ and the define of FWHM of a Gaussian distribution in Eq. (C4), we obtain the standard deviation $\sigma_y$ of $y$ as

$$\sigma_y = \frac{4A(ln4)^{3/2}}{\sqrt{\pi}\mathrm{FWHM}^3}\sigma_n^2. \quad \text{(C5)}$$

$\sigma_y$ is just the NAMNM.

*Contact author: chenxi@apm.ac.cn
†Contact author: wangjin@apm.ac.cn
‡Contact author: mszhan@apm.ac.cn

*Contact author: chenxi@apm.ac.cn
†Contact author: wangjin@apm.ac.cn
‡Contact author: mszhan@apm.ac.cn

*Contact author: chenxi@apm.ac.cn
†Contact author: wangjin@apm.ac.cn
‡Contact author: mszhan@apm.ac.cn